\documentclass{PoS}

\usepackage{amsmath}

\title{Simulating quantum field theory \\with a quantum computer}

\ShortTitle{Simulating quantum field theory with a quantum computer}

\author{\speaker{John Preskill}
\\
        Institute for Quantum Information and Matter \\
	Walter Burke Institute for Theoretical Physics \\
	California Institute of Technology, Pasadena CA 91125, USA\\
        E-mail: \email{preskill@caltech.edu}}

\abstract{
Forthcoming exascale digital computers will further advance our knowledge of quantum chromodynamics, but formidable challenges will remain. In particular, Euclidean Monte Carlo methods are not well suited for studying real-time evolution in hadronic collisions, or the properties of hadronic matter at nonzero temperature and chemical potential. Digital computers may never be able to achieve accurate simulations of such phenomena in QCD and other strongly-coupled field theories; quantum computers will do so eventually, though I'm not sure when. Progress toward quantum simulation of quantum field theory will require the collaborative efforts of quantumists and field theorists, and
though the physics payoff may still be far away, it's worthwhile to get started now. Today's research can hasten the arrival of a new era in which quantum simulation fuels rapid progress in fundamental physics. 

}

\FullConference{The 36th Annual International Symposium on Lattice Field Theory - LATTICE2018\\
		22-28 July, 2018\\
		Michigan State University, East Lansing, Michigan, USA.}

\begin{document}

\section{Introduction}

My talk at Lattice 2018 had two main parts. In the first part I commented on the near-term prospects for useful applications of quantum computing. In the second part, I assessed the prospects for advancing fundamental physics through simulations of quantum field theory using quantum computers and quantum simulators. In this article based on that talk, I will de-emphasize the first topic, because I have already written up my thoughts about that in \cite{preskill-nisq}. Instead, I'll insert some general remarks about quantum field theory which might be helpful for quantum computationalists who lack familiarity with this subject.  



Along the way I'll tell you about some of the things I've done with my collaborators (Stephen Jordan, Hari Krovi, and Keith Lee), some of the things we still hope to do, and why we are doing these things \cite{jlp1,jlp2,jlp3,jklp}. This is a high-level overview, eschewing technical details. I have tried to make it accessible to a broader audience than the attendees of Lattice 2018.

\section{Quantum simulation of quantum field theory: Why do it?}

\subsection{The Quantum Church-Turing Thesis}
Physics underlies computer science, because computation is a physical process. The fundamental question about computers is: What computations can be performed efficiently, and what computations are intractable? This is a question about the laws of physics and the consequences of those laws. 

Though we don't know how to prove it from first principles, we have good reasons to believe that there are computational problems which are too hard to solve using digital computers based on classical physics, yet can be solved by computers which exploit quantum phenomena such as interference and entanglement. This is one of deepest distinctions ever made between classical and quantum physics, and it poses the compelling challenge to identify those problems which are classically hard but quantumly easy. 

The Church-Turing Thesis asserts that any physically reasonable model of computation can be simulated by a (classical) Turing machine. This statement is still widely believed. What has been upended by the discovery of fast quantum algorithms is a stronger statement regarding computational \textit{complexity}, sometimes called the \textit{Strong} Church-Turing Thesis, which asserts that any physically reasonable of computation can be \textit{efficiently} simulated by a classical Turing machine. This statement is probably wrong, and should be replaced by an updated version, the \textit{Quantum} Church-Turing Thesis. Informally, this revised thesis asserts that the quantum computers which we plan to build and operate in the coming decades will be capable of efficiently simulating \textit{any} process that occurs in Nature. 

Is the Quantum Church-Turing Thesis true? We don't know, but either a Yes or No answer is pleasing. If the answer is Yes (and if as we suspect, the classical version of the Strong Church Turing Thesis is false), then the quantum computer will become an indispensable tool for exploring fundamental physics. If the answer is No, that's even more exciting; it means our current concept of a quantum computer doesn't fully capture the computational power that's encoded in the laws of Nature.

\subsection{The future of computational quantum chromodynamics}


All the known fundamental interactions in Nature, with the conspicuous exception of gravitation, are encompassed by the Standard Model of particle physics. Physicists work hard to compute the consequences of the Standard Model, and to compare these predictions with experimental results, with great success. The most daunting computational challenge has been the study of quantum chromodynamics (QCD), the quantum field theory underlying the structure and interactions of \textit{hadrons}, strongly interacting composite particles such as  mesons, protons, neutrons, and atomic nuclei.

The Hamiltonian of QCD is precisely known, but the Schr\"odinger equation for this Hamiltonian is too hard to solve analytically, because hadronic matter is subject to large quantum fluctuations. Fortunately, some predictions of QCD can be extracted through numerical Monte Carlo simulations. The study of \textit{Lattice QCD} has been a major thrust in computational physics for decades, and many impressive results have been achieved. In particular the masses of hadrons have been precisely computed and found to match their experimentally observed values; furthermore, many phenomenologically relevant matrix elements of local operators have been accurately estimated. 

With greater processing power, such static quantities can be computed with improved statistics. In a few years, exascale Monte Carlo simulations of lattice QCD may provide adequate precision for many typical physics applications \cite{brower}. Where should lattice QCD go from there?

In fact, the Monte Carlo simulations have serious limitations. Even with further advances in digital computing technology, many quantities of interest to physicists will remain out of reach because of a fundamental obstacle known as the \textit{sign problem}. In particular, we would like to be able to simulate the real time \textit{dynamics} of high-energy collisions between hadrons, computing from first principles the distribution of outgoing particles produced in such collisions. This information is needed to guide searches for new physics beyond the Standard Model in experiments performed using particle accelerators. We would also like to know the properties of nuclear matter at high density, pressure, and temperature. This information is needed for studies of supernovae and neutron stars, and for understanding the behavior of hadronic matter in the early history of the universe. Yet these tasks seem to be beyond the reach of digital computation because of the sign problem.

These tasks are difficult because they involve simulations of \textit{highly entangled matter}, which cannot be succinctly described in ordinary classical language, and hence cannot be accurately encoded in conventional digital computers. Quantum computers \textit{can} simulate highly entangled systems of many particles, and hence solve the sign problem. Eventually (I'm not sure when), quantum computers will be essential for advancing our understanding of QCD. 

\subsection{Beyond QCD}

Aside from QCD, quantum simulation will also allow us to explore the properties of other strongly-coupled quantum field theories, which might be important for formulating models of physics beyond the Standard Model. 
In addition, simulations of quantum field theory provide a stepping stone to simulations of quantum gravity, which (at least in the case of quantum gravity in anti-de Sitter space) can be formulated in terms of a quantum field theory living on the boundary of spacetime \cite{maldacena}. 
Beyond providing quantitative results,  both thinking about and actually performing simulations of quantum field theory and quantum gravity may well lead to unanticipated new conceptual insights.

\section{The potential of quantum computing}

\subsection{Why we think quantum computing is powerful}
We have at least three good reasons for thinking that quantum computers have capabilities surpassing what classical computers can do. (1) We know of problems that are believed to be hard for classical computers, but for which quantum algorithms have been discovered that could solve these problems easily. The best known example is the problem of finding the prime factors of a large composite integer \cite{shor}. (2) Arguments based on complexity theory show (under reasonable assumptions) that a quantum computer can efficiently sample from probability distributions that we can't sample from efficiently by any classical means \cite{bremner,harrow-montenaro}. (3)  After many decades of effort by physicists to find better ways to simulate quantum systems, we still don't know how to efficiently simulate a quantum computer using a classical digital computer.

\subsection{Why quantum computing is hard}
But building and operating a large-scale quantum computer is a formidable task. Why is it so difficult? The core of the problem stems from a fundamental feature of the quantum world --- we cannot observe a quantum system without producing an uncontrollable disturbance in the system. That means that if we want to use a quantum system to store and reliably process information, then we need to keep that system nearly perfectly isolated from the outside world. 

Eventually we expect to be able to protect quantum systems and scale up quantum computers using the principle of quantum error correction \cite{gottesman}. The essential idea  is that if we want to protect a quantum system from damage then we should encode it in a very highly entangled state; then the environment, interacting with parts of the system one at a time, cannot glimpse the encoded information and therefore can't damage it. 
Unfortunately, there is a significant overhead cost for doing quantum error correction,
 so reliable quantum computers using quantum error correction are not likely to be available very soon.

\subsection{The NISQ era unfolds}



Though truly scalable fault-tolerant quantum computing is still a rather distant dream, we are now entering a pivotal new era in quantum technology. 
I thought we should have a name to describe this impending new era, so I made up a word: {\textit{NISQ} \cite{preskill-nisq}. It stands for \textit{Noisy Intermediate-Scale Quantum}. Here ``intermediate scale'' refers to the size of quantum computers which will be available in the next few years, with a number of qubits ranging from 50 to a few hundred;
these devices are too large to be simulated by brute force using the most powerful existing digital supercomputers.
``Noisy'' emphasizes that we'll have imperfect control over those qubits. Noise will limit the power of NISQ technology, because if we attempt to execute too large a quantum circuit (say, one with more than 1000 gates --- that is, 1000 fundamental two-qubit operations), then the signal in the final readout is likely to be overwhelmed by the noise.


Physicists are excited about this NISQ technology, which gives us new tools for exploring the physics of many entangled particles.  
We also hope that NISQ devices will have valuable practical applications, but we're not sure about that. 
I do think that quantum computers will have transformative effects on society eventually, but these may still be decades away. We just don't know how long it's going to take.

\subsection{Quantum simulation}
Of the applications for quantum computers that we can currently foresee, the most important is likely to be 
studying the properties of highly entangled many-particle quantum systems. 
We think that simulating such ``strongly correlated'' matter is a difficult computational problem, because many very good physicists have tried to solve it for decades and have not succeeded. 


Classical computers are especially bad at simulating \textit{quantum dynamics} --- that is, predicting how a highly-entangled quantum state will change with time. That is the natural arena where quantum computers seem to have a clear advantage over classical ones \cite{feynman}. Perhaps we will learn interesting things about quantum dynamics using NISQ technology in the relatively near future, using noisy devices with of order 100 qubits.

\subsection{Digital vs. analog quantum simulators}

When we speak of an \textit{analog quantum simulator} we mean a system with many qubits whose dynamics resembles the dynamics of a model system we are trying to study and understand. In contrast, a \textit{digital quantum simulator} is a gate-based universal quantum computer which can be used to simulate any physical system of interest when suitably programmed, and can also be used for other purposes.

Analog quantum simulation has been a very vibrant area of research for the past 15 years, while digital quantum simulation with general purpose circuit-based quantum computers is just now getting started. 
Analog quantum simulators have been getting notably more sophisticated, and are already being employed to study quantum dynamics in regimes which may be beyond the reach of classical simulators \cite{lukin,monroe}. 

We can anticipate, though, that analog quantum simulators will eventually become obsolete. Because they are hard to control, they will be surpassed some day by digital quantum simulators, which can be firmly controlled using quantum error correction. Nevertheless, because of the hefty overhead cost of quantum error correction, the reign of the analog quantum simulator may persist for many years. 
When seeking near-term applications of quantum technology, we should not overlook the potential power of analog quantum simulators. 

\subsection{The daunting climb to scalability}


Solving really hard problems (like factoring numbers which are thousands of bits long) using fault-tolerant quantum computing is not likely to happen for a while, because of the large number of physical qubits needed. To run algorithms involving thousands of protected qubits we may need a number of physical qubits which is in the millions, or more \cite{fowler}.  That's a very large leap from where we will be for the next few years, with of order a hundred qubits. Crossing the ``quantum chasm,'' from hundreds of physical qubits to millions of physical qubits (and beyond), is going to take some time, but we'll get there eventually.

Throughout the NISQ era it will remain important to strive for lower gate error rates in the various quantum platforms. With more accurate quantum gates, quantum computers operating without error correction will be able to go further by executing larger circuits. Furthermore, better gates will lower the overhead cost of doing quantum error correction when we eventually move on to fault-tolerant quantum computing in the future. 
To the extent possible, we should also build noise resilience into our quantum 
algorithms. 


\section{Quantum field theory 
}
\subsection{Some history}
Quantum field theory was first invented soon after the arrival of quantum mechanics; 
Dirac, Heisenberg, Jordan, Pauli, Born  and other pioneers of quantum mechanics also contributed to early formulations of quantum field theory. One of the first things they noticed is that quantizing the free electromagnetic field yields photons; that insight provided a firmer foundation for the concept of light quanta, which until then had been a more heuristic idea. 

Because the coupling strength of photons to electrons and other charged particles (the fine-structure constant $\alpha \approx 1 /137$) is weak, it's sensible to use perturbation theory to work out the consequences of the quantum electrodynamics, and this works pretty well for computations to the first nontrivial order in $\alpha$. But confusion reigned for 20 years because when one attempts to improve the accuracy of such computations by going to higher order in perturbation theory, infinities are encountered which seemed at first to have no clear physical interpretation. That situation changed abruptly in the late 1940s, when Schwinger, Feynman, Tomonaga, and Dyson realized that one can handle these infinities successfully by thinking operationally about what is being computed. When comparing the result of a computation with experiments, we should express our answers in terms of other experimentally accessible quantities, such as the mass or charge of the electron; then the infinities conveniently cancel. 
The moral is: if you ask the theory a sensible physical question it will provide a sensible answer; however, it took a lot of combinatoric agility
to verify that this procedure (called \textit{renormalization}) really works. 

The first efforts to turn quantum field theory into a rigorous mathematical subject occurred in the 1950s, when Wightman in particular formulated a set of axioms which define what we mean by a relativistic quantum field theory \cite{wightman}. The subject took a significant step forward around 1970 largely through the work of Ken Wilson, who taught us to think of quantum field theory as a kind of second-order phase transition \cite{wilson}. The Standard Model of particle physics was also formulated in the 1970s, and still stands as our best description of the strong and electroweak interactions after decades of thorough vetting in high-energy-physics experiments.

\subsection{Universality: Why Ken Wilson is our hero}
Ken Wilson is a hero to me and many others because he provided a satisfying answer to the question: What is quantum field theory? (His was not the first answer, nor was it the complete and final answer, but nevertheless it transformed our understanding of the subject.) Wilson understood more deeply than his predecessors the meaning of renomalization. 

The problem of infinities arises because quantum field theory 
involves an infinite number of degrees of freedom per unit volume; the fields have modes of arbitrarily short wavelength. Even for a finite volume of space, no matter how small, simulating this infinite system exactly would be impossible with any finite machine. But the key point about renormalization is that if you are interested in computing things that can be measured at long distances or low energy (in what physicists call ``the infrared''), there is limited sensitivity of that long-distance physics to the underlying physics at very short distances (``the ultraviolet''). 

You might think that the long-distance physics should depend in a very complicated way on the short-distance physics that defines the theory, and that's true in a way. But something wonderful happens. All of the short-distance details can be absorbed into a small number of ``renormalized parameters'' needed to specify the long-distance phenomena. We call this miracle ``universality'' to convey that many models with different properties at short distances can all give rise to the same predictions about the physics at long distances.  

Universality makes quantum field theory useful in condensed matter physics, where we typically study macroscopic phenomena which are insensitive to the microsopic properties at the atomic distance scale. And it is also used in fundamental physics to describe elementary particles and their interactions. For this latter purpose the quantum field theories of interest are ``relativistic'' --- this means the same laws of physics govern all observers who move at constant velocity relative to one another, so there is no preferred inertial frame of reference. 

\subsection{Effective field theory}

Universality is really what makes it possible for us to do physics at all. If we had to understand physics at the Planck scale, at $10^{-35}$ meters, to compute the properties of the hydrogen atom ($10^{-10}$ meters) or the physics that's going on at the Large Hadron Collider ($10^{-18}$ meters), then we'd really be stuck. Fortunately, that is not necessary. We can do physics one scale at a time, understanding a given energy scale in detail, while still remaining largely ignorant about what's going on at still higher energies. 

By attacking problems one energy scale at a time we can make progress, but universality is really a two-edged sword.  Since low-energy phenomena do not depend very strongly on physics at shorter distances, it is hard to learn about very-short-distance physics in experiments which are feasible using today's particle accelerators. This limitation frustrates the high-energy physicists. 

From today's post-Wilson viewpoint, the notion that a quantum field theory accommodates an infinite number of degrees of freedom per unit volume is really a convenient fiction, an idealization rather than a precise description of reality. The Standard Model accurately describes almost all known physics (excluding gravity), but we modestly call it an ``effective'' field theory. The word ``effective'' reminds us of the theory's limited scope, that we expect it to fail eventually if we push it toward and higher and higher energy. 

An effective theory has an energy scale associated with it, what we call the theory's renormalization scale. As that scale changes, the effective theory should change, too, if it is to continue to provide an accurate description of the physics at that scale. This scale-dependence of the theory is called ``renormalization-group flow.'' 
Given an effective theory, a theory which makes sense at shorter distances and flows to that effective theory at longer distances is called the ``ultraviolet completion'' of the effective theory. If an ultraviolet completion of the effective theory exists, because of universality it need not be unique. That's why, to find out how physics behaves at shorter distances, we need to do experiments at higher energies. 



Sometimes instead of an ``ultraviolet completion'' we speak of a ``continuum limit.'' Following Wilson, we make sense of the quantum field theory by replacing continuous space by a spatial lattice with a nonzero lattice spacing $a$, where the Hamiltonian couples neighboring lattice sites. This theory defined on the lattice, with a finite number of lattice sites per unit of physical volume, is said to be a ``regulated'' version of the continuum theory, where the lattice spacing can be regarded as the ``short-distance cutoff,'' the shortest distance scale to which the theory is applicable. Once the theory is regulated, it has a finite number of degrees of freedom per unit volume, rather than an infinite number, which makes it easier to define, understand, and study. Furthermore, once regulated, the theory can be simulated using methods which are well known to computational scientists. 

Taking the ``continuum limit'' in which the lattice spacing $a$ shrinks to zero is essentially equivalent to characterizing what the physics looks like at distance scales which are arbitrarily large compared to the lattice spacing. This limit may fail to exist, or rather, it might be that the theory becomes a boring noninteracting (``free'') theory in this limit. Correspondingly, in that case the interacting effective theory cannot be completed in the ultraviolet.

Wilson's insights largely flowed from his thinking, in the 1960s, about how to simulate a quantum field theory on a digital computer. It is reasonable to anticipate that we will gain further insights  from thinking about, and actually doing, simulations of quantum field theories on quantum computers. Or we might have an even more ambitious goal --- perhaps thinking about, and actually doing, simulations of quantum gravity on quantum computers can bring us closer to answering the vexing question: What is string theory? It may be noteworthy that the time it took from the first formulations of quantum field theory (ca. 1927), to Wilson's  insights about the nature of the subject (ca. 1970),  is comparable to the span of time since string theory was first proposed as a quantum theory of gravity (ca. 1974) to today. In the study of quantum gravity we're still waiting for our Wilson to come along.

One apologetic note: I have heavily used the word ``theory'' in this discussion. Physics readers are not likely to be perturbed by overuse of this word, but for specialists in other fields, like computer science, it can be disorienting. I have not attempted to explain what I mean by a ``theory,'' and there may not be unanimous agreement about that even among physicists.  Suffice it to say that, thanks to universality, we can identify broad classes of microscopic model Hamiltonians which describe similar long-distance physics, encoded in a small number of adjustable renormalized parameters, and we usually regard two Hamiltonians in the same class as belonging to the same ``theory.'' In many cases, just the spatial dimensionality and the symmetry of the Hamiltonian are enough data for specifying the ``theory'' in question. 

\subsection{About rigor}
For mathematically inclined readers, I'll make a comment about rigor. One can rigorously define what a relativistic quantum field theory is --- that was point of the Wightman axioms --- and one can prove nontrivial theorems that follow from these axioms \cite{wightman}. In the early days this program was nonconstructive; the only known examples of theories satisfying the axions where ``free'' theories of noninteracting particles, which are not very interesting. Starting in the 1970s there were notable successes at rigorously constructing nontrivial interacting theories that satisfy the axioms \cite{jaffe}, but that success has mostly been limited to what we call \textit{superrenormalizable} theories, which have especially weak sensitivity to the short-distance physics. It remains unsettled whether asymptotically free quantum chromodynamics (QCD) exists as a theory that satisfies the Wightman axioms, though we physicists have no particular reason to doubt it. 

With rare exceptions, quantum field theory in more than $D=4$ spacetime dimensions  isn't a very interesting subject because only free theories exist . The case of a self-coupled scalar theory in $D=4$ is marginal; we don't think it exists as an interacting theory in the continuum satisfying all the Wightman axioms, but it is still potentially interesting to simulate its behavior at distances exceeding the short-distance cutoff by a reasonably large factor. 

When we analyze classical or quantum algorithms for simulating quantum field theory, we sometimes need to settle for heuristic arguments. For example, in estimating how the error in our simulations scales with the lattice spacing, we might make use of perturbation theory in ways that we can't fully justify \cite{jlp1,jlp2,jlp3}.  For now that's the best we can do. Our goal is for the analysis to be precise when possible, and nonrigorous only when necessary. 

By the way, while there is a rigorous version of relativistic quantum field theory, and a set of axioms specifying properties which any such theory should obey, we don't yet have that for string theory. A famous conjecture asserts that quantum gravity in negatively curved anti-de Sitter (AdS) spacetime is precisely equivalent (or ``dual'') to a conformally invariant quantum field theory defined on the boundary of the spacetime \cite{maldacena}. Though there is much evidence in favor, this conjecture is a bit dissatisfying at present because we have no alternative way of defining quantum gravity in AdS aside from saying it is the thing which is dual to the field theory!

\section{Quantum simulation of quantum field theory: What is it?}


\subsection{What problem does the algorithm solve?} 
When we simulate quantum field theory, what kinds of problems might we want to solve? There are many options.

We might, for example, 
study \textit{scattering processes}. Given some description of an initial state such as particles that are about to collide with one another, we can sample accurately from the final states produced by collisions. We might want to compute the \textit{vacuum-to-vacuum persistence amplitude}. If classical sources are coupled to local observables, where the sources turn on and off in some way that depends on position in space and time, we can calculate the amplitude for the ground state of the theory to be preserved. This amplitude can be related to \textit{correlation functions} of the local observables. We can also calculate such correlation functions directly, determining the matrix element between initial and final ground states for a string of local observables inserted at specified points in spacetime. Such correlators can be related to bulk \textit{transport properties} like thermal conductivity or viscosity. Or we can compute transport properties by directly simulating the bulk dynamics. To study phenomena of interest, it is often fruitful to ask: How would you investigate these properties by doing an experiment? Your answer may guide the formulation of an algorithm modeled on that potential experiment. 

All these simulations of dynamics are hard to do with classical computers. But no ``sign problem'' prevents us from doing the simulation if we have a quantum computer.

\subsection{Real time vs. imaginary time}
Experts on lattice gauge theory should be reminded that when we speak of algorithms for simulating quantum systems on a quantum computer we usually have in mind simulating quantum evolution in real time. Today's lattice experts prefer (classical) simulations in imaginary time, in Euclidean rather than Lorentzian spacetime. We quantumists wish we could do the same, but we don't know how to simulate imaginary time evolution with a quantum computer. 

That's too bad, because imaginary time evolution has advantages for some purposes like efficiently preparing ground states of quantum systems. But it's not really so bad, because Nature evolves in real time, too, and our ultimate goal is to simulate how Nature behaves! Furthermore, the simulation of real-time evolution for highly entangled many-particle quantum systems, including quantum field theories, is believed to be a hard problem classically, yet quantumly tractable. Euclidean computations are useful for computing static properties, like the mass spectrum and matrix elements of local operators, but are ill suited for studying dynamical phenomena. That's where the quantum computer should have a huge advantage. 

In quantum simulations we work with Hamiltonians, not with a manifestly covariant formulation of the theory like an action formulation. 
Typically we pick some inertial frame, and simulate the Hamiltonian evolution in that frame. If we are careful we can extract results which do not depend on that choice of reference frame. In any case, our goal might be to simulate an experiment as observed in a particular frame, so it makes sense to simulate the dynamics as seen in that frame. 

The Hamiltonians that arise in physics are typically \textit{local}. When the quantum algorithm experts say a Hamiltonian $H$ is local, they usually mean that $H$ can be expressed as a sum of terms such that each term couples together only a constant number of degrees of freedom, rather than a number of degrees of freedom which increases with the size of the system. When physicists say $H$ is local, they usually have in mind a stronger property, sometimes called \textit{geometric locality} (in a specified number of spatial dimensions). For example, a Hamiltonian is geometrically local in one dimension if the degrees of freedom can be arranged on a line, and each term in the Hamiltonian couples only degrees of freedom located inside an interval of constant length. 

Quantum field theories are geometrically local in this sense; if the theory is defined on a spatial lattice, only fields at neighboring lattice sites are coupled. Local Hamiltonians are easier to simulate than arbitrary Hamiltonians, but the simulation task can still be quite challenging!

\subsection{Prototypical task}
Here's a typical task that we might perform in a quantum simulation. (1) Prepare the initial state whose evolution we want to study in our quantum computer, such as an incoming scattering state of two or more particles, if we are trying to study a scattering process. (2) Evolve the state forward in time using the Hamiltonian, in the particular reference frame that we've chosen, for some specified time interval. (3) Measure an observable, for example by simulating the measurement performed by a particle detector in an idealized laboratory. 

The goal of this procedure is to sample accurately from the probability distribution of possible outcomes for that final measurement. If the simulation is imperfect, then we quantify the error in the simulation as the distance (in some suitable norm) between the probability distribution we sample from in practice, and the ideal distribution for the noiseless system.

In analyzing the algorithm, we determine what computational resources would suffice to achieve a specified error. The resources we typically care about are  the number of qubits and the number of gates, perhaps also the circuit depth if the computation is highly parallelizable.   This resource cost depends on many features of the input to the problem. These include the simulated physical volume, the evolution time, the number of particles involved, the total energy of the process, and the desired error. The cost also depends on properties of the Hamiltonian, in particular on particle masses and interaction strengths. 

If we use a classical computer to perform this task, a brute force simulation would require resources scaling exponential with the system size.
If we use a quantum computer instead, we hope at the very least that  the resource cost scales polynomially with all of these input parameters, and we have verified this scaling for scattering problems involving scalar particles and fermions \cite{jlp1,jlp2,jlp3}. 
If the resources scale polylogarithmically rather than polynomially with the input parameters, then we are even happier. Constant factors are also important, especially if we wish to assess whether the algorithm might be practical using relatively near-term devices.



\subsection{Preparing the initial state}
To get the simulation started, we need to load the initial state into our quantum computer. That might not be so easy. For some local Hamiltonians, even preparing the ground state of the Hamiltonian is computationally hard, too hard for a quantum computer to solve the problem efficiently. As we've known for decades, just finding the lowest energy state of a classical Ising-like spin glass in two dimensions is NP-hard \cite{barahona}, as hard as any problem whose solution can be checked efficiently using a classical computer. For quantum local Hamiltonians, preparing the ground state is even harder: it is \textit{QMA-hard}, as hard as any problem whose solution can be checked efficiently using a quantum computer \cite{kitaev}, and QMA-hard instances can occur even in one dimension \cite{gottesman-irani}.

That sounds scary, but we don't need to be too discouraged if our goal is to simulate Nature. States existing in Nature are not ones which are NP-hard or QMA-hard to prepare. (If they were, that would be a deep and very surprising discovery about the history of the universe.) This remark applies to ground states, and also to equilibrium Gibbs states at nonzero temperature and chemical potential, or to states far from equilibrium. For a spin glass, for example, preparing a low-temperature Gibbs state might be hard, but that won't prevent us from simulating the behavior of the spin glass observed in an actual experiment performed in a reasonable amount of time; if it takes an exponentially long time for the system to reach thermal equilibrium, then equilibrium behavior will not be observed in the experiment. 

With admitted hubris, we declare that if Nature has managed to prepare the initial state efficiently, then we should be able to do the same in our simulation. But how? No method is foolproof, but there are a few tricks at our disposal.

One is the \textit{adiabatic method}. For example, suppose we would like to prepare the ground state of a massive interacting scalar field theory. If we turn off the coupling constant, the theory becomes free; then its ground state is Gaussian, and hence easy to prepare. Then we simulate the evolution of the state as the coupling constant slowly increases to the desired nonzero value. Thanks to the quantum adiabatic theorem, the system is likely to stay in its ground state during the evolution if the coupling changes slowly, on a  time scale determined by the \textit{mass gap}, the mass of the lightest particle. This method may fail if a quantum phase transition is encountered during the excursion of the coupling, where the mass gap goes to zero. Even in that case we might succeed when simulating a finite volume, since the finite-size system has a nonzero energy gap even at the critical point, but the very small gap due to the finite volume means that the coupling must change especially slowly while crossing the critical point, which makes the method much less efficient. 

Of course, the vacuum does not evolve, but we can adapt the adiabatic method to prepare more interesting states. For example, after the vacuum of the interacting theory is prepared, we can simulate carefully shaped pulses coupled to the fields which excite the creation of localized particle wave packets with specified (approximate) momentum and position \cite{jklp}. Alternatively, we can prepare wave packets in the free theory (which is fairly easy), and then ``tether'' the wave packets to prevent them from propagating or spreading during the adiabatic ramping up of the coupling \cite{jlp1,jlp2}.

An alternative approach is to leverage the success of classical \textit{tensor-network methods}, which work especially well in one spatial dimension, for preparing low-energy states of many-body systems. 
For example, classical methods can be used to find an efficient description of the ground state \cite{dmrg}, or of the incoming state for a scattering process \cite{verstraete-tangent}, which can be then be compiled as a quantum circuit to initialize a quantum computation \cite{jordan}.


\subsection{How to regulate}
How should we regulate a quantum field theory and turn it into an object that we know how to simulate? There are a lot of options, and it is not immediately obvious which way is best. 

There are potential advantages to working in momentum space, since transforming to momentum space diagonalizes the Hamiltonian of a free translation-invariant theory, and furthermore there are powerful tools for studying renormalization-group flow in momentum space. However, for simulating interacting theories the position-space formulation of the theory is more natural, because the interactions among fields are spatially local; therefore working in the position basis tends to make the simulation more efficient.

In our algorithms for simulating scalar field theories, we introduce a spatial lattice, with lattice spacing $a$. The lattice is an artifice introduced for convenience, and our goal is to extract physical properties at distance scales much larger than $a$, which are relatively insensitive to the precise structure of the theory at scale $a$.

Formally, fields are operator-valued distributions \cite{wightman}. More informally, a scalar field $\phi(x)$ is an unbounded quantum continuous variable at each point $x$ in space. For the theory to be simulable, we replace $\phi(x)$ at each lattice site by a discrete variable with a finite number $N$ of mutually orthogonal eigenstates. There is a conjugate variable $\pi(x)$, the field momentum at the lattice site $x$, which is related to $\phi(x)$ by Fourier transforming. Here I mean, not the Fourier transform with respect to the spatial variable $x$, but rather the quantum Fourier transform applied to the $N$-dimensional Hilbert space residing at each lattice site $x$. For $N=2^n$, the quantum state of the field at a site can be encoded in $n$ qubits.

This truncation of $\phi(x)$ and $\pi(x)$, as well as the nonzero lattice spacing $a$, are sources of error. If we want to study a process with energy $E$, then as $E$ increases we need to make the lattice spacing $a$ smaller in physical units, and we also need to increase the dimension $N$ of the Hilbert space at each site, to keep the error fixed. This means that the number of qubits needed for the simulation of a fixed physical volume increases with $E$. 

There may be more clever ways of regulating that would improve the efficiency of the simulation. Alternatives have been proposed \cite{verstraete-cmps,verstraete-cmera}, but I don't know of persuasive evidence that an alternative method would significantly improve the resource cost of simulations that address physics questions of interest. 


\subsection{Example: $\phi^4$ theory in $D$ spacetime dimensions}

To be a bit more concrete, consider the Hamiltonian for a self-interacting scalar field in $D$ spacetime dimensions ($D-1$ spatial dimensions):

\begin{eqnarray}\label{eq:phi4-ham}
H &=& H_\pi + H_\phi, \quad \textrm{where}\nonumber\\
H_\pi &=& \sum_{x\in\Lambda} a^{D-1} \left(\frac{1}{2}\pi(x)^2\right),\nonumber\\
H_\phi &=& \sum_{x\in\Lambda} a^{D-1} \left(\frac{1}{2}\nabla \phi(x)^2 + \frac{m_0^2}{2} \phi(x)^2 + \frac{\lambda_0}{4!}\phi(x)^4\right).
\end{eqnarray}
Here the lattice site $x$ is summed over a cubic lattice $\Lambda$, and the gradient term is defined by 
\begin{equation}
\nabla \phi(x)^2 = \sum_{j=1}^{D-1} \left(\frac{\phi(x+ e_j) - \phi(x-e_j)}{2a}\right)^2,
\end{equation}
where the $x+e_j$ denotes the nearest neighbor of site $x$ in the $j$ direction, and $a$ is the lattice spacing. The conjugate variables $\phi(x)$ and $\pi(x)$ obey the canonical commutation relations
\begin{equation}
\left[\phi(x),\phi(y)\right] = 0, \quad \left[\pi(x),\pi(y)\right] = 0 , \quad \left[\phi(x),\pi(y)\right] = ia^{-(D-1)} \delta(x,y). 
\end{equation}

If the coefficient $\lambda_0$ of the $\phi^4$ term vanishes, then the Hamiltonian density is a quadratic function of the $\phi$ and $\pi$ variables. In that case the theory is Gaussian; it is easy to diagonalize the Hamiltonian and solve the theory exactly --- it describes massive noninteracting particles. When $\lambda_0$ is nonzero these particles interact and the physics is more interesting. The dimensionless number which characterizes the strength of the these interactions is $\lambda_0 /m_0^{4-D}$. However, the parameters $m_0$ and $\lambda_0$ are the \textit{bare} parameters appearing in the lattice Hamiltonian; $m_0$ may have a much different value than the actual physical mass of the particle described by this theory, and $\lambda_0 m_0^{4-D}$ may be much different than the dimensionless coupling constant which characterizes the actual interaction strength for particles with energy small compared to $a^{-1}$.

Note that we have written the Hamiltonian density in eq.(\ref{eq:phi4-ham}) as a sum of two terms --- one is diagonal in the $\pi(x)$ basis, while the other is diagonal in the $\phi(x)$ basis. We can switch back and forth between these two bases by Fourier transforming the field variables at each site \cite{jlp1,jlp2,klco2}. (I emphasize again that this Fourier transform should not be confused with the spatial Fourier transform.) Decomposing the Hamiltonian this way is handy when we simulate time evolution. If we choose a small time step $\epsilon$, we can express $\exp(-i\epsilon H)$ as $\exp(-i\epsilon H_\pi )\exp(-i\epsilon H_\phi)$ making a small $O(\epsilon^2)$ error (and furthermore the error can be systematically suppressed using higher-order Suzuki-Trotter formulae \cite{suzuki}). It is easy to simulate time evolution governed by the diagonal Hamiltonian $H_\pi$ or $H_\phi$. Thus we can evolve the system using the (field) Fourier transform to alternate back and forth between the $\pi$ and $\phi$ bases, and applying a diagonal evolution operator in each small time step.

In the Hamiltonian formulation, introducing a lattice breaks Lorentz invariance badly. Therefore, we need to carefully tune the bare parameters in the lattice Hamiltonian to closely approximate the Lorentz-invariant theory we wish to study. In addition, to accurately describe continuum physics we need the physical mass $m$ of the lightest particle in the theory to be small compared to $1/a$, where $a$ is the lattice spacing, and fine tuning of the bare Hamiltonian is needed for this purpose as well. (That we need to tune close to a second-order phase transition for a scalar particle to be light compared to the ultraviolet cutoff was the reason for expecting the discovery of new physics at the Large Hadron Collider at energy scales not far above the Higgs boson mass.)

\subsection{Sources of error}
Suppose we simulate the high-energy scattering of two single-particle wave packets in the self-coupled scalar field theory. As described in \S 5.4, we might do this by first preparing wave packets of the free theory, and then ``dressing'' the wave packets by adiabatically turning on the coupling constant, while keeping the wave packets tethered to prevent propagation and spreading. Once the wave packets are fully dressed, we untether them, and allow them to collide; as described in \S 5.6, we can simulate time evolution using Trotter product formulas, while applying the quantum Fourier transform to switch back and forth repeatedly between the $\pi(x)$ basis and the $\phi(x)$ basis. 
Finally we sample the final state by measuring suitable observables. We might do this by adiabatically turning off the coupling and then using \textit{quantum phase estimation} to measure occupation numbers of the modes of the free theory. (In a typical high-energy collision, many particles are produced.) Here we take for granted that there are no avoided level crossings obstructing the adiabatic method;  potentially these could arise because of a quantum phase transition encountered during the excursion of the coupling, or because the lightest particles in the interacting theory are bound states rather than dressed versions of the  single-particle states of the free theory.

What potential sources of error should we worry about when we follow this protocol? Some of these are already familiar from experience with Euclidean lattice computations. The simulation has a nonzero lattice spacing, and we need to extrapolate to the limit of zero lattice spacing to capture the continuum physics accurately. We simulate only a finite spatial volume, and so need to extrapolate to the limit of infinite volume. In addition, the fields and conjugate momenta at each lattice site are unbounded continuous variables, which we approximate by retaining only a finite number of bits of precision. Aside from all that, we approximate continuous time evolution with a finite quantum circuit, introducing an error which depends on the size of the time step. Finally, when we use the adiabatic method for state preparation, diabatic particle production can be a serious contribution to the error if the Hamiltonian changes too quickly. The combined effect of all these sources of error determines how the computational resources scale with the error and with the properties of the incoming scattering state. Finally, all of the errors described so far arise even when the quantum circuit used in the simulation is executed perfectly. In a realistic simulation, noise in the execution of quantum gates will be a further troublesome source of error.



To be a bit more concrete, consider the case of the scalar field theory with a $\phi^4$ interaction in 2+1 spacetime dimensions \cite{jlp1,jlp2}. This theory is superrenormalizable, so we know it has a continuum limit and a relatively well behaved perturbation expansion. A perturbative analysis indicates that the error $\epsilon$ in the simulation scales with the lattice spacing $a$ as $\epsilon = O(a^2)$. The number of sites needed to simulate a physical spatial volume  $V$ is $V/a^2$; therefore the number of qubits $n$ needed scales like $n=O\left(V/\epsilon\right)$ for a fixed spatial volume, assuming we keep fixed the number of qubits per lattice site. For processes at relatively low energy, the most expensive part of the algorithm is the initial preparation of the Gaussian state for the free theory. This is a matrix algebra task, which can be achieved with $G = O(n^{2.373}) = O\left( (V/\epsilon)^{2.373})\right)$ gates. 

This simulation algorithm has not been carefully costed --- in our initial work we only studied the asymptotic scaling of the cost. But rough estimates indicate that simulating dynamics on, say, a $32 \times 32$ lattice would already require thousands of logical qubits and millions of logical gates. That's discouraging. 

By being more clever we might do better. For example, with a careful choice for the lattice Hamiltonian the leading dependence of the error on the lattice spacing $a$ can cancel, improving how $G$ scales with $\epsilon$. Furthermore, our naive estimate for the Gaussian preparation step assumes we use a general procedure which would apply to any Gaussian state of $n$ qubits, while the ground state of the free theory has a special property -- translation invariance -- which reduces the cost. 
We also hope and expect that there are more efficient protocols than those described in \cite{jlp1,jlp2}, based on different ideas.



One may also wonder how the number of gates $G$ scales with the total energy $E$ of the simulated scattering process. 
In our analysis of the algorithm \cite{jlp1,jlp2}, we concluded that $G = O\left(E^6\right)$, though this estimate is probably too pessimistic. First, as we increase the energy we choose a smaller lattice spacing in order to resolve shorter-wavelength physics; this accounts for two factors of $E$. Second, we choose a smaller Trotter step size to maintain a fixed Trotter error; that's another factor of $E$. 

But the dominant effect is that as $E$ increases more computational steps are needed during the adiabatic dressing of the single-particle wave packet states.
During the slow turning on of the coupling constant, the Hamiltonian remains translation invariant, so that momentum is conserved, but diabatic jumps across an energy gap can occur if the Hamiltonian changes quickly enough. For example, the process in which a single particle with energy $E$and mass $m$ splits into three particles moving almost collinearly has an energy gap which scales like $m^2/E$ when $E/m$ is large. (It is also possible for diabatic effects to produce a pair of particles anywhere in the sample volume, but that process is less likely because the energy gap $2m$ is larger.) Such unwanted particle production is adequately suppressed if the adiabatic dressing takes longer (hence requiring more gates) as $E$ increases, accounting for the three additional powers of $E$. 

Most likely a less conservative analysis would find more favorable scaling of $G$ with $E$, but it may be difficult to do accurate and convincing estimates of the computational cost of such quantum simulation algorithms until we are able to run them on actual quantum devices. 

\subsection{Simulations of one-dimensional physics}
Studies of quantum field theory in one spatial dimension will require more modest resources than simulations in higher dimensions, so it is natural to explore the opportunities for answering well-motivated physics questions in simulations of one-dimensional quantum field theory.
What quantum simulations of one-dimensional dynamics might surpass what can be done classically? 
Particularly powerful classical tools exist for simulating one-dimensional physics, especially for systems with an energy gap. Therefore, in defining the task achieved by a quantum simulation, we should strike a balance between choosing a task which is as easy as possible for our quantum simulator, yet sufficiently hard for the best existing classical simulators that the quantum simulator might have a clear advantage. The key is to make sure that the quantum states to be simulated become so highly entangled that classical methods fail. 

Time evolution in a one-dimensional systems can be simulated classically using, for example, the time-evolving block decimation (TEBD) method or the time-dependent variational principle (TDVP) method
\cite{verstraete-tangent}. 
In both cases, one uses a matrix-product state (MPS) approximation to the quantum state, which is continually updated as the state evolves.
Accuracy is limited by the maximal feasible bond dimension of the tensors, and the approximation fails badly if the entanglement of the state grows beyond what the maximal bond dimension can accommodate. 

If we consider a high-energy collision between two particles in one dimension, how entangled are the outgoing particles? A crude picture is that the collision creates a ``fireball'' and that many approximately thermal particles are emitted as the fireball cools and dissipates. If the temperature of the fireball is $O(1)$, then the number of particles produced, and the thermodynamic entropy of the outgoing particles, scales linearly with the total energy of the process. 
If the incoming state is pure, then so is the outgoing state, with the momentum balanced between right-moving and left-moving particles. 
The thermodynamic entropy of the particles moving in one direction is really entanglement entropy, arising from their correlations with the particles moving in the opposite direction. According to this naive picture, then, we would expect the entanglement arising from the collision to increase linearly with the total energy $E$ and the total number of particles $N$ produced. The bond dimension needed to accurately describe a state with bipartite entanglement entropy $S \approx N$ would scale like $e^N$. This may already become classically intractable for $N\approx 10$. In contrast, the quantum simulation remains efficient even when the state becomes highly entangled. 

Alternatively, we might simulate a ``quench'' in which the Hamiltonian suddenly changes. After the quench, a state which is initially slightly entangled becomes more highly entangled as it evolves \cite{cardy}. The bond dimension needed in an MPS description grows, and classical simulation may eventually become infeasible if the energy density is nonzero.

For either the scattering process or the quench, the initial state in the simulation need not be highly entangled, so that it admits a succinct classical MPS description which can be found using a \textit{classical} optimization algorithm. Once found, that MPS description can be compiled as a quantum circuit for preparing the initial state, at a lower cost in quantum gates than a fully quantum protocol for the state preparation. Even in the case of a highly relativistic collision that produces many particles, the initial state of the incoming particles is only modestly entangled, and therefore has an accurate MPS description with a reasonable bond dimension. 

More generally, because of the limitations of NISQ technology, we may anticipate that the early advances in computational physics enabled by quantum devices will make use of hybrid classical-quantum algorithms, in which the power of a classical supercomputer is somehow enhanced by a quantum co-processor. Using a classical computer to find the MPS description of the initial state in a quantum simulation is one example of how we can exploit the complementary capabilities of the classical and quantum computers.





It is also noteworthy that, even for non-relativistic particles at weak coupling in one spatial dimension, multiple scattering events can build up enough entanglement for classical simulation to be hard. In fact, simulating many particles which scatter many times is BQP-hard \cite{jklp}; if we could solve that problem efficiently with a classical computer (using resources scaling polynomially with the number of particles and the number of scattering events), then the classical computer could efficiently solve any problem that a quantum computer can solve efficiently. We don't think that's possible, and therefore have good reason to believe that simulating a scalar field theory in one dimension (even at weak coupling) is a hard problem for classical computers, yet easy for quantum computers. 

\subsection{The challenge of gauge theories}

The focus of this meeting is the exploration of quantum chromodynamics using computational tools. For reasons already discussed, quantum computers will eventually be used for this purpose, but it's not clear when. Even for a relatively small $32^3$ lattice, with a standard encoding we might need millions of logical qubits, taking into account the multiple colors and flavors. 

A more efficient encoding of quantum states might reduce this cost substantially. Perhaps that is possible, in view of the huge gauge redundancy that arises using brute-force encodings. In a one-dimensional gauge theory, where ``photons'' or ``gluons'' are nondynamical,  the spatial position of charges, together with boundary conditions, suffice to determine the gauge fields; therefore we can solve the Gauss law constraint exactly, eliminating the gauge fields to obtain a geometrically nonlocal Hamiltonian involving only charged matter degrees of freedom \cite{zohar3}. This trick has been exploited successfully to study both static and dynamical properties in one dimension, using both classical and quantum simulators \cite{muschik,blatt-zoller}. So far, though, there have not been attempts to do quantum simulations of dynamical processes that produce highly entangled states that are hard to simulate classically. Perhaps this will be possible soon, for one-dimensional gauge theories. Unfortunately, this method of integrating out gauge fields does not work, at least not in the same way, in higher dimensional theories with dynamical gluons or photons. It's important to explore other ways of reducing the gauge redundancy \cite{klco1}, with the goal of formulating more efficient simulation protocols for dynamical phenomena in QCD.

Meanwhile, there are also ingenious suggestions for using analog simulation platforms like ultracold atoms in optical lattices to simulate gauge theories in one or more spatial dimensions \cite{reznik,zohar4,wiese-zoller1,wiese-zoller2,oberthaler1}. These proposed experimental protocols are certainly challenging, but still may be more powerful than digital quantum simulations of gauge theories for the time being. It's not yet clear, however, to what extent these methods can achieve sufficient accuracy to extract valuable information surpassing what can be attained using classical simulation tools. 

\section{Challenges and opportunities in quantum simulation of quantum field theory}

\subsection{Where are we now?}

The study of quantum algorithms for simulating quantum field theory is still in its very early stages. Some achievements so far include:


$\bullet$ The first crude (and probably overly pessimistic) estimates of asymptotic resource scaling (number of qubits and gates) for simulations of high energy scattering in scalar and Yukawa theories \cite{jlp1,jlp2,jlp3}. The resources scale polynomially with total energy, particle number, and accuracy, but are nonetheless sobering. 
\smallskip

$\bullet$ A demonstration that simulating a scalar quantum field theory in one dimension is BQP hard, even at weak coupling, in scenarios where many particles scatter many times \cite{jklp}. This argument bolsters the case that quantum simulations of dynamics using quantum computers can surpass classical methods for simulating quantum field theory. 
\smallskip

$\bullet$ Successful applications of tensor network methods in classical simulations of one-dimensional gauge theories with massive fermions \cite{pichler,banuls2,verstraete-schwinger1,banuls4}. Both static properties and dynamical processes (like the breaking of an electric flux tube due to spontaneous creation of electrically charged fermion pairs) have been studied. 
\smallskip

$\bullet$ Few-site quantum simulations of 1D QED using trapped ions and superconducting circuits \cite{muschik,blatt-zoller,klco1}. 
\smallskip

$\bullet$ Proposals for analog quantum simulations using ultracold atoms \cite{reznik,zohar4,wiese-zoller1,wiese-zoller2,oberthaler1}.
\smallskip

$\bullet$ Proposals for digital quantum simulation of nonabelian gauge theories in higher dimensions \cite{yamamoto,zohar1}.

\subsection{Where should we go?}

What should we be thinking about next?

$\bullet$ It would be useful to do more serious costing of the algorithms proposed so far, and to seek improvements. More clever methods of regularizing and renormalization group improvement of lattice Hamiltonians might help a lot.
\smallskip

$\bullet$ We need to think through what physics insights could be drawn from simulations with NISQ devices, perhaps by exploiting hybrid quantum-classical methods. Studies of dynamics are potentially promising, but it may be crucial to improve the noise resilience of digital quantum simulation protocols, if we hope to run useful simulations using gates with relatively high noise and without using quantum error-correcting codes. 
\smallskip

$\bullet$ For simulations of gauge theories beyond one spatial dimension, reducing gauge redundancy in encoding of quantum states could improve resource requirements substantially. 
\smallskip

$\bullet$ We need better methods and analysis for dealing with systems that have massless particles (like the photon). Long-wavelength massless particles will be copiously produced during adiabatic changes in the Hamiltonian; we should design our algorithms to address physics questions that are insensitive to these spurious soft quanta. 
\smallskip

$\bullet$ Chiral fermions pose another important challenge.  The standard model is chiral; that is, for both quarks and leptons, the massless left-handed and right-handed fermions carry different gauge charges. Yet existing methods for formulating fermion theories on the lattice always yield nonchiral theories --- \textit{e.g.}, if we try to introduce left-handed fermions with a specified charge we also get unwanted right-handed particles with the same charge. We've been facing this problem for over 40 years, but there is still no accepted method for regulating a chiral theory. That's embarrassing. 

This long-standing problem may be nearing a resolution, guided in part by recent insights regarding symmetry-protected topological phases of matter. Two old ideas are: (1) To realize a $D$-dimensional chiral theory on the lattice, we can introduce an extra spatial dimension, so that the left-handed and right-handed fermions live on two different $D$-dimensional edges of a $(D+1)$-dimensional bulk \cite{kaplan}. (2) To realize a $D$-dimensional chiral theory on the lattice, we can introduce strong interactions for the express purpose of removing the unwanted right-handed fermions (by giving them large masses) while preserving the massless left-handed fermions \cite{eichten}. It seems likely that (1) and (2) together work more effectively than either (1) or (2) by itself \cite{wen}. That's because separating the two edges with a higher-dimensional bulk makes it easier to apply the strong interactions to one chirality without affecting the other. 

The efficacy of this method still needs to be demonstrated convincingly, but if it works that will settle the longstanding open question whether quantum field theories with chiral fermions really exist, and will also open the door for classical and quantum studies of the rich dynamics of strongly-coupled chiral gauge theories, with potential applications to physics beyond the standard model. 


\smallskip

$\bullet$ Realizing supersymmetry on the lattice will also create new opportunities for informative studies of dynamics. There has been progress on this problem in the Euclidean framework \cite{schaich}, but so far realizations of supersymmetry in the Hamiltonian setting have received less attention. 
\smallskip

$\bullet$ Also of great interest is the dynamical behavior of conformal field theories, with or without supersymmetry. Part of my own not-so-secret motivation for pursuing quantum simulations is that I hope that some day we will enrich our understanding of holographic duality through the simulations of holographic CFTs. Possibly this will be achieved not just through the kind of brute-force simulation considered in this article, but also using quite different methods. It may be noteworthy, in particular, that the conformal bootstrap program allows us to extract valuable information about CFTs by approximately solving semidefinte programs, a task that might be performed much faster with quantum computers than classical ones \cite{bao}. 
\smallskip

$\bullet$ In any case, we should continue the quest for alternative paradigms which will enrich our conceptual grasp of quantum field theory  as well as suggesting new computational methods.
I firmly believe that the circle of ideas relating quantum chaos, quantum information, quantum entanglement, and emergent geometry offers great opportunities for major steps forward in theoretical physics; quantum simulation of strongly-coupled dynamics may help to fuel further progress. 

\smallskip
Quantum simulation of quantum field theory will be a long term project, and it may be a while before transformative physics results can be achieved. Ultimately, though, I think such simulations will be seen as an especially important consequence of the development of quantum technology. To make progress, the quantum algorithm experts and the lattice field theorists will need to join forces and pool their expertise. There is much to be learned as the subject advances over the decades ahead. So let's get going ---  this is going to be fun!


\section*{Acknowledgments}

My remarks here have been influenced by many colleagues, most of all my collaborators Stephen Jordan, Hari Kvoi, and Keith Lee. I'm also grateful for illuminating discussions with (among others) Alex Buser, David Kaplan, Alexei Kitaev, Natalie Klco, Junyu Liu, Benni Reznik,  Burak \c{S}ahino\u{g}lu, Martin Savage, Frank Verstraete, and Erez Zohar.
My work is supported by ARO, DOE, IARPA, NSF, and the Simons Foundation. The Institute for Quantum Information and Matter (IQIM) is an NSF Physics Frontiers Center.

\end{document}